\documentclass[%
 reprint,
 amsmath,amssymb,
pra,
]{revtex4-2}

\usepackage{graphicx}
\usepackage{dcolumn}
\usepackage{bm}
\usepackage{xcolor}

\begin{document}

\preprint{APS/123-QED}

\title{Theory of filter-induced modulation instability in driven passive optical resonators}


\author{Auro M. Perego$^1$}
\email{a.perego1@aston.ac.uk}
\author{Arnaud Mussot$^{2}$ }
\author{Matteo Conforti$^2$}
\email{matteo.conforti@univ-lille.fr}
\affiliation{%
$^{1}$Aston Institute of Photonic Technologies, Aston University, Birmingham, B4 7ET, UK\\
$^{2}$Univ. Lille, CNRS, UMR 8523-PhLAM-Physique des Lasers Atomes et Mol\'ecules, F-59000 Lille, France
}%

\date{\today}

\begin{abstract}
We present the theory of modulation instability induced by spectrally dependent losses (optical filters) in passive driven nonlinear fiber ring resonators. Starting from an Ikeda map description of the propagation equation and boundary conditions, we derive a mean field model - a generalised Lugiato-Lefever equation - which reproduces with great accuracy the predictions of the map. The effects on instability gain and comb generation of the different control parameters such as dispersion, cavity detuning, filter spectral position and bandwidth are discussed. %
%

\end{abstract}

\maketitle


\section{Introduction}
Modulation instability (MI) is an ubiquitous phenomenon occurring in various fields of nonlinear physics consisting in the exponential amplification of spectral sidebands which results in a modulation of a powerful and originally constant amplitude wave \cite{Zakharov}. Besides leading to the destabilization of nonlinear waves in a vast range of contexts including fluids dynamics \cite{Zakharov1968}, optics \cite{Bespalov}, plasmas \cite{Ichikawa1973} and Bose-Einstein condensates \cite{Carr2004}, MI is deeply connected to solitons dynamics and is the initiating mechanism for pattern formation process too \cite{Cross1993}.
It is customary to understand MI as a synchronization process between a powerful wave corresponding to the unstable homogeneous state of the system and detuned spectral sidebands whose amplitude is very small in the initial stage.
The waves synchronization process is determined by a phase mismatch parameter, which accounts for physical effects describing dephasing between different waves. Waves for which mismatch is close to zero synchronize with the powerful homogeneous mode and energy transfer from the powerful wave to the sidebands occurs, causing to the exponential growth of the latter.
A vast range of modulation instabilities have been studied in the literature. 
The most paradigmatic MI example is definitely the Benjamin-Feir instability, originally studied in fluid dynamics \cite{BF,Bespalov,Zakharov} and later in nonlinear fibre optics \cite{Tai1986}. It can be understood as  a nonlinear four waves interaction enabled by the interplay of the cubic nonlinearity and anomalous group velocity dispersion. It can be mathematically described in the framework of the nonlinear Schr\"odinger equation (NLSE).
Another celebrated MI is the Turing instability \cite{Turing1952}, which can arise in passive driven optical Kerr resonators described by the Lugiato-Lefever equation (LLE) \cite{Lugiato,Haelterman92,Oppo2009} where, in addition to dispersion and nonlinearity, the detuning between pump and cavity resonance may enable MI in otherwise stable regime (in the Benjamin-Feir framework).

The longitudinal modulation of parameters such as group velocity dispersion \cite{Smith1996,Conforti2014,Conforti2016,Copie2016,Copie2017,Mussot2018} and nonlinearity coefficients \cite{Abdullaev,Staliunas,Matera1993} in NLSE and LLE may support instabilities as well, which are analogous to the parametric (Faraday) instability \cite{Coullet1994}. In this case quasi-phase matching conditions describing synchronization between spectral sidebands and homogeneous mode can be obtained allowing precise estimation of the amplified frequencies, which determines the parametric resonances.

A further different class of MI relies on homogeneous or periodic action of spectrally dependent losses.
The periodic case results in a dissipative parametric (Faraday) instability, which proves relevant for achieving high repetition rate mode-locking in lasers \cite{Perego1,Perego2}.
Homogeneously distributed frequency dependent losses,  can result in counterintuitive amplification of damped modes themselves. This happens if losses act in unbalanced fashion on two sidebands waves whose frequencies are symmetrically located with respect to a powerful input one, which in absence of losses  would be stable \cite{PeregoL} (a fortiori for symmetric losses \cite{Karlsson1995}). This case has been first analysed by Tanemura and co-authors in an optical fibre \cite{Tanemura}, and also subsequently described by other authors  using coupled mode theory (non-Hermitian phase-matching)  \cite{NH,NH2}.

Presence of spectrally dependent losses can also indirectly induce phase-matching by modifying the mismatch parameter of the system via the phase profile naturally associated to dissipation by Kramers-Kronig relations.  
Examples of the latter are provided in studies of the resonant dispersion MI \cite{Murdoch,Kalithasan_2010,Mirtchev1,Mirtchev2} where the phase in the vicinity of an atomic resonance can contribute to phase-matching.
Recently an example of dissipation induced modulation instability has been reported in a driven passive ring fiber resonator with intracavity spectral filter \cite{Bessin}. In that work, gain-through-filtering enabled by filter phase modification of the cavity mismatch was suggested as a novel method for optical frequency comb with tuneable repetition rate generation in normal dispersion regime.

The aim of this article is to report a comprehensive theoretical description of filter-induced modulation instabilities in passive cavities. We first review and expand the theory of filter induced instability based on the Ikeda map approach originally described in \cite{Bessin}.  We derive a mean-field generalised Lugiato-Lefever equation and develop a linear stability analysis of this model. The mean-field approximation is showed to permit a simpler, yet accurate, description with respect to the Ikeda map.
Finally, we discuss the dependence of the instability gain on various system parameters, 
and the generation of pulse trains and frequency combs. 

\section{The Ikeda map}
The evolution of a light pulse in a fiber ring resonator can be modelled by the following coupled equations, referred to as Ikeda map \cite{Ikeda1979,Ikeda1980,Ikeda1982} (see also \cite{Coen1997,Conforti2016,Conforti2017}) :
%
\begin{align}
\label{nls} i\frac{\partial A_n}{\partial z}-\frac{\beta_2}{2}\frac{\partial^2 A_n}{\partial t^2}+\gamma|A_n|^2A_n=0,\;\; 0<z<L,\\
\label{bc} A_{n+1}(z=0,t)=\theta E_{IN}+\rho e^{i\phi_0}A_n(z=L,t).
\end{align}
Here $A(z,t)$ represents the slowly varying envelope of the electric field, normalised in such a way that $|A|^2$ has the dimensions of a power. The coordinate $t$ is the retarded time and the spatial coordinate $z$  measures the position inside a fiber ring cavity of length $L$.  $\gamma$ is the Kerr nonlinearity coefficient, $\beta_2=\partial^2_\omega \beta|_{\omega=\omega_p}$ is the group velocity dispersion coefficient at the pump wavelength, with $\beta$ the propagation constant of the fiber mode and $n$ an integer counting the number of cavity round trips.
All the losses (except the filter induced ones) are lumped in $\rho$, so that $1-\rho^2$ measures the total power loss per roundtrip. $\phi_0=[\beta(\omega_p) L \mod 2\pi]$ is the linear phase shift per roundtrip modulo $2\pi$ (the cavity detuning is $\delta=-\phi_0$) and $\theta$ is the transmission coefficient of the coupler for the pump amplitude $E_{IN}$. 
For simplicity, we neglect higher order dispersion terms, as it allows to accurately model most of realistic configurations, but they can be introduced in a straightforward fashion if needed. Note that only even order of dispersion contribute to the modulation instability gain. 
The filter located at the position $z=z_F$ acts in the following way:
\begin{align}
A_n(z_F^+,t)&=h(t)\star A_n(z_F^-,t),\\
\hat A_n(z_F^+,\omega)&=H(\omega)\,\hat A_n(z_F^-,\omega),
\end{align}
where $h(t)$ is the filter impulse response (causality imposes $h(t)=0$ if $t<0$), $\star$ denotes convolution and $H(\omega)=\hat h (\omega)=\int_{-\infty}^{+\infty} h(t)\exp[i\omega t]dt$ is the filter transfer function. 
The filter is assumed to be placed just before the coupler ($z_F=L$), hence the boundary conditions and filter can be conveniently combined in the single equation:
\begin{equation}\label{newBC}
A_{n+1}(z=0,t)=\theta E_{IN}+\rho e^{i\phi_0}h(t)\star A_n(z=L,t),
\end{equation}
which has the following equivalent in the frequency domain:
\begin{equation}\label{newBCfreq}
\hat A_{n+1}(z=0,\omega)=\theta E_{IN}\delta(\omega)+\rho e^{i\phi_0}H(\omega)\hat A_n(z=L,\omega),
\end{equation}
where $\delta(\omega)$ is the Dirac delta function.
\subsection{The filter}
As a consequence of physical causality, the real and imaginary parts of the filter transfer function $H(\omega)$ are related by the Kramers-Kronig (KK) relations \cite{Lucarini}. This implies that 
the presence of losses entails a corresponding phase shift. KK relations lead to a similar connection between the magnitude and the phase, which is known as Bode or Bayard-Bode (BB) gain-phase (or magnitude-phase) relation \cite{bode}:
\begin{eqnarray}\label{Bode}
\psi(\omega)={\mathcal{H}\{F(\omega)\}=\mathcal{H}\{\log|H(\omega)|\}},
\end{eqnarray}
where we have defined  
\begin{equation}\label{filt}
H(\omega)=e^{F(\omega)+i\psi(\omega)},
\end{equation}
$\displaystyle \mathcal{H}\{f(x)\}=\frac{1}{\pi}\mathrm{p.v.}\int_{-\infty}^{+\infty}\frac{f(y)dy}{x-y}$ is the Hilbert transform \cite{Bracewell,Hilbert,notehilbert}, p.v. denoting the principal value of the integral. 
The advantage of Bode relation is that it is straightforward to experimentally measure the amplitude of the response, while it is tricky to access real or imaginary parts.
While KK is an equality, Bode's relation is an inequality, which can under-estimate the phase response. BB coincides with KK only if $\log|H(\omega)|$ is analytic and $H(\omega)\neq 0$ in the upper-half complex-$\omega$ plane. Response functions having these additional properties are called minimum-phase, and in this study we will consider only this kind of filter for which Eq. (\ref{Bode}) holds \cite{note_delay}.

\begin{figure}[htb]
\includegraphics[width=0.45\textwidth]{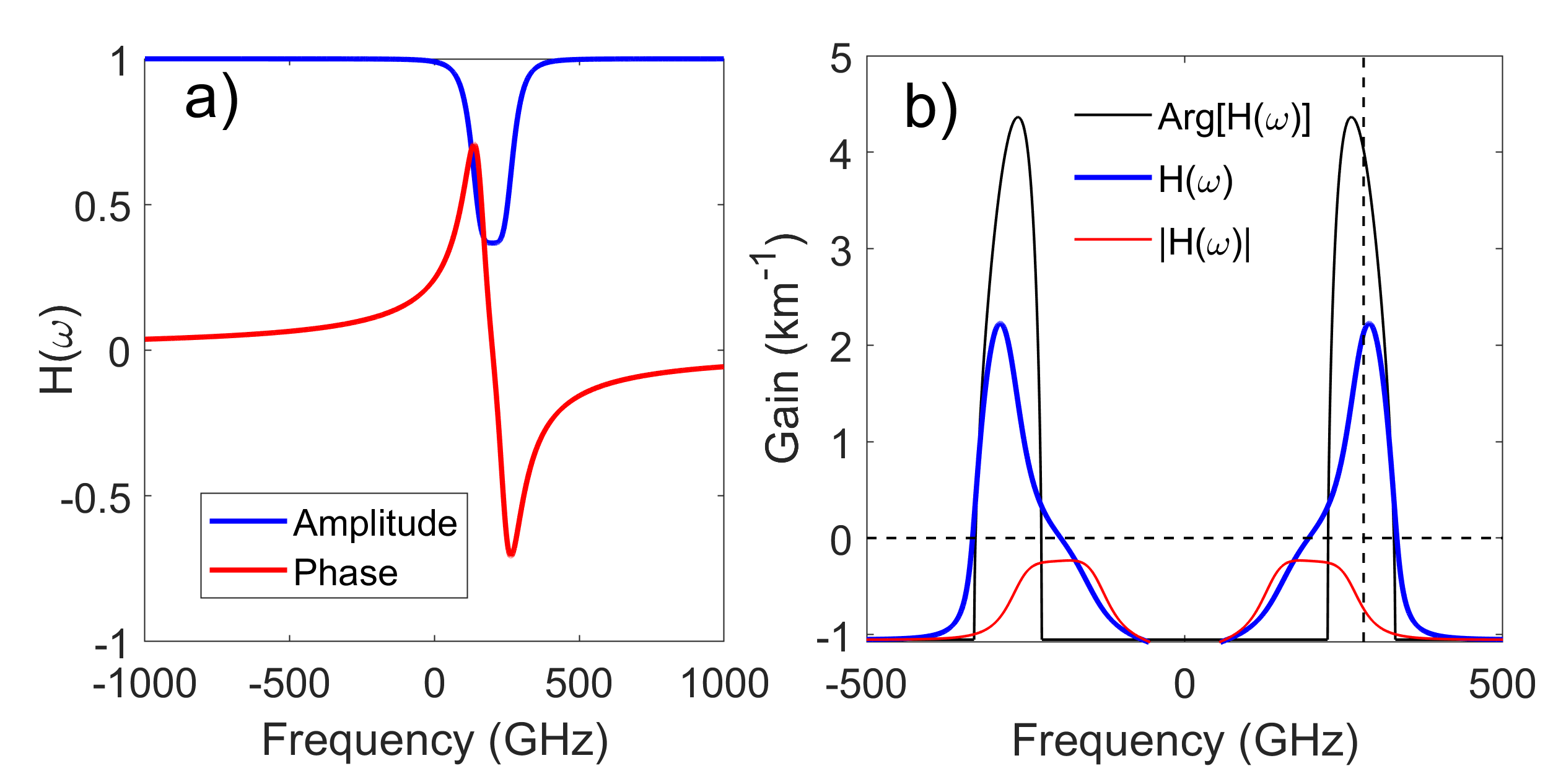}
\caption{(a) Amplitude and phase of the filter described by Eqs. (\ref{powerF},\ref{phaseF}). (b) Instability gain $g_{MAP}(\omega)$ calculated from Eq. (\ref{gain_map}) revealing the effect of magnitude and phase of the filter transfer function. The following parameters have been used: $\beta_2=0.5$ ps$^2$/km , $\gamma=2.5$ W$^{-1}$km$^{-1}$, $L=100$ m, $\rho=\sqrt{0.9}$, $\theta=\sqrt{0.1}$, $\phi_0=-\psi(0)$ (zero global cavity detuning), $\omega_\text{f}=200\cdot 2\pi $ rad/ns, $b=-1$, $a=400$  rad/ns, intracavity power $P=1.18$ W,  input power $P_\text{IN}=1$ W. The vertical dashed line indicates the phase-matching frequency calculated from Eq. (\ref{PM1}). }\label{figure1} 
\end{figure}

In the following we will consider a higher order Lorentzian filter for which the Hilbert transform can be calculated analytically \cite{Hilbert}
\begin{eqnarray}
\label{powerF} F(\omega)&=&b\frac{a^4}{(\omega-\omega_f)^4+a^4},\\ \label{phaseF}\psi(\omega)&=&
ba^4\frac{(\omega-\omega_f)\left[(\omega-\omega_f)^2+a^2\right]}{\sqrt{2}[(\omega-\omega_f)^4+a^4]}
\end{eqnarray}
where $a$ is related to filter bandwidth (in rad/s) and $b<0$ is a non-dimensional number which controls the filter strength, i.e. the maximum attenuation of the filter. The half-width of the filter at half-attenuation can be easily computed as $\Delta\omega_{HWHM}=a\sqrt[4]{b/\ln[(1+e^b)/2]-1}\approx a(1-b/8)$  for small $b$.
In addition to provide a simple and elegant analytical description of a causal filter, Eqs. (\ref{powerF},\ref{phaseF}) also provide a good approximation of the transfer function of apodised fiber Bragg gratings (FBG), as the one used in \cite{Bessin}. It is worth noting that FBGs used in transmission are always minimum phase \cite{Lenz1998,Poladian1997}, which makes our analysis rather general. For other kinds of filters with arbitrary dissipation profile, for which no analytical expression is known, the corresponding phase can be calculated numerically \cite{Lucarini,Hilbert}.
An example of the transfer function of the filter described by Eqs. (\ref{powerF},\ref{phaseF}) is shown in Fig. \ref{figure1}(a).
%
%
\subsection{Steady states}
We search for the stationary, continuous wave (CW) 
field inside the fibre as
$$A_n(z,t)=\overline A e^{i\gamma P z}, \;\; P=|\overline A|^2.$$
The relation between the field circulating into the cavity and the pump is (for complex field and power)
\begin{align}
\overline A&=\frac{\theta}{1-\rho e^{i\phi} H(0)  }E_\text{IN},\label{staz}\\
P&=\frac{\theta^2}{1+\rho^2|H(0)|^2-2\rho|H(0)| \cos(\phi+ \psi(0)) }P_\text{IN},\label{staz2}
\end{align}
where the total phase shift imposed by the cavity  is $\phi=\phi_0 +\gamma P L$ while $E_\text{IN}$ and $P_\text{IN}$ denote the amplitude and the power of the pump field respectively. From Eq. (\ref{staz2}) it appears that the magnitude of the filter at the pump frequency $|H(0)|$ acts as an additional loss which multiplies the coupler induced losses, whereas the phase $\psi(0)$ gives an additional phase shift.

\subsection{Linear stability analysis}
In order to study the stability of the steady state defined by  Eq. (\ref{staz}) we consider the following perturbation
$$A_n(z,t)=[\sqrt P +\eta_n(z,t)]e^{i\gamma P z},\;\;|\eta_n|\ll \sqrt P. $$
For simplicity we have assumed the intracavity field to be real (the steady state phase does not affect the results of the stability analysis). 
By linearisation we obtain the equation governing the evolution of the perturbations:
$$i\frac{\partial \eta_n}{\partial z}-\frac{\beta_2}{2}\frac{\partial^2\eta_n}{\partial t^2}+\gamma P (\eta_n +\eta_n^*)=0. $$
We split perturbations in real and imaginary parts $\eta_n=a_n+ib_n$ ($a_n,b_n\in \mathbb{R}$), to get the following system describing the evolution of the perturbations' spectra $\hat a=\hat a(z,\omega)$, $\hat b=\hat b(z,\omega)$ :
\begin{align}
\label{az} \frac{\partial{\hat a_n}}{\partial z}&=-\frac{\beta_2\omega^2}{2}\hat b_n,\\ 
\label{bz} \frac{\partial{\hat b_n}}{\partial z}&=\left (\frac{\beta_2\omega^2}{2}+2\gamma P\right)\hat a_n.
\end{align}

The solution of the system Eqs. (\ref{az}-\ref{bz}) from $z=0$ to $z=L$ gives the perturbations' spectra after one pass in the fibre as (dependence on frequency $\omega$ is omitted):
\begin{equation}\label{X}
\left[
\begin{array}{ c }
\hat a_n(L) \\
\hat b_n(L)
\end{array}
\right] =
\left[
\begin{array}{ c c}
\cos(kL)  & -\frac{\beta\omega^2}{2k}\sin(kL)\\
\frac{2k}{\beta\omega^2}\sin(kL) & \cos(kL)
\end{array}
\right]
\left[
\begin{array}{ c }
\hat a_n(0) \\
\hat b_n(0)
\end{array}
\right],
\end{equation}
where $k(\omega)=\sqrt{\frac{\beta_2\omega^2}{2}\left(\frac{\beta_2\omega^2}{2}+2\gamma P\right)}$ is the wave-number of the small harmonic perturbations which propagate on top of the stationary field.

From Eqs. (\ref{newBC}-\ref{newBCfreq}) it follows that the combined action of the filter and the coupler on the perturbations can be written as follows:
\begin{align}\label{BC}
\nonumber&\left[
\begin{array}{ c }
\hat a_{n+1}(0) \\
\hat b_{n+1}(0)
\end{array}
\right] =\\
&\rho\left[
\begin{array}{ c c}
\cos\phi  & -\sin \phi\\
\sin \phi & \cos \phi
\end{array}
\right]
\left[
\begin{array}{ c c}
H_\text{e}(\omega)  & -H_\text{o}(\omega)\\
H_\text{o}(\omega) & H_\text{e}(\omega)
\end{array}
\right]
\left[
\begin{array}{ c }
\hat a_n(L) \\
\hat b_n(L)
\end{array}
\right],
\end{align}
where we have defined the even and odd part of the transfer function as
\begin{align*}
H_\text{e}(\omega)=\mathcal F\{\mathrm{Re}[h(t)]\}=\frac{H(\omega)+H^*(-\omega)}{2},\\
H_\text{o}(\omega)=\mathcal F\{\mathrm{Im}[h(t)]\}=\frac{H(\omega)-H^*(-\omega)}{2i}.
\end{align*}
Note that the coupler and filter matrices commute, as expected intuitively: for the stability analysis it doesn't matter if the filter is placed just before or just after the coupler.

By combining Eqs. (\ref{X}-\ref{BC}) we get the total effects accumulated by the perturbations over one roundtrip as
\begin{equation}
\left[
\begin{array}{ c }
\hat a_{n+1}(0) \\
\hat b_{n+1}(0)
\end{array}
\right] =
M
\left[
\begin{array}{ c }
\hat a_n(0) \\
\hat b_n(0)
\end{array}
\right].
\end{equation}
The eigenvalues of the matrix $M$ reads as
\begin{equation}\label{eigs}
\lambda_{1,2}=\frac{\Delta}{2}\pm\sqrt{\frac{\Delta^2}{4}-W}
\end{equation}
where
\begin{align}\label{W}
W&=\rho^2\left(H_\text{e}(\omega)^2+H_\text{o}(\omega)^2\right),\\
\nonumber \Delta&=\rho\bigg[2\cos(kL)\left(H_\text{e}(\omega) \cos \phi - H_\text{o}(\omega) \sin \phi\right)\\\label{delta}
&-\frac{\beta_2\omega^2+2\gamma P}{k} \sin(kL)\left(H_\text{o}(\omega) \cos \phi +H_\text{e}(\omega) \sin \phi\right) \bigg].
\end{align}
Whenever $|\lambda_{1,2}|>1$ the CW solution Eq. (\ref{staz}) is unstable and the perturbation power grows as $\exp[g_{MAP}(\omega)z]$, where we have defined the MI gain as:
\begin{equation}\label{gain_map}
g_{MAP}(\omega)=\frac{2}{L}\ln \max\{|\lambda_1|,|\lambda_2|\}.
\end{equation}

%
\subsection{Approximations and phase matching condition}
Equations (\ref{eigs}-\ref{gain_map}) give the exact parametric gain, however they do not allow for a straightforward physical interpretation. We hence proceed to obtain an approximated formula, which holds valid when the MI process derives mainly from the filter phase. 
Indeed, by exploring the parameters' space, we have noted that the position of the unstable bands is mainly fixed by the filter phase. 
Figure \ref{figure1} shows the relative impact of the filter amplitude, phase and both combined on the instability gain spectrum. The red curve, accounting for only $|H(\omega)|$, shows that the threshold of instability is not reached. The most unstable band mimics the shape of the filter response, as expected for gain-through-loss mechanism \cite{PeregoL}. 
The black curve accounts only for the filter phase:  even if it overestimates the gain, it gives a reasonable prediction of the frequency of the unstable bands. In order to predict position of the unstable bands, we then assume the following form for the filter transfer function (unitary modulus)
\begin{equation}
H(\omega)=\exp[i\psi(\omega)].
\end{equation}

The even and odd part of the filter transfer function read as
\begin{align}
H_\text{e}(\omega)=e^{i\psi_\text{o}(\omega)}\cos[\psi_\text{e}(\omega)],\\
H_\text{o}(\omega)=e^{i\psi_\text{o}(\omega)}\sin[\psi_\text{e}(\omega)],
\end{align}
where the even and odd part of the filter phase are defined as
\begin{align}
\psi_\text{e}(\omega)=\frac{\psi(\omega)+\psi(-\omega)}{2},\\
\psi_\text{o}(\omega)=\frac{\psi(\omega)-\psi(-\omega)}{2}.
\end{align}
The assumption of unitary modulus, permits to greatly simplify Eqs. (\ref{W}) as follows
\begin{align}\label{Ws}
W&=\rho^2e^{i2\psi_\text{o}},\\
\nonumber \Delta&=\rho e^{i\psi_\text{o}}\bigg[2\cos(kL)\cos(\phi+\psi_\text{e})\\
&-\frac{\beta_2\omega^2+2\gamma P}{k} \sin(kL)\sin(\phi+\psi_\text{e})\bigg]
\triangleq e^{i\psi_\text{o}}\tilde\Delta,
\end{align}
which gives the following expression for the eigenvalues:
\begin{equation}\label{eigss}
\lambda_{1,2}=e^{i\psi_\text{o}}\bigg[\frac{\tilde\Delta}{2}\pm\sqrt{\frac{\tilde\Delta^2}{4}-\rho^2}\bigg].
\end{equation}
A part from the exponential factor, Eq. (\ref{eigss}) has been obtained before for the description of a standard cavity (i.e. without filter) \cite{Coen1997,Conforti2016}. The exponential factor does not change the modulus of the eigenvalues, hence it does not affect the gain.
We have instability if $|\tilde \Delta|>1+\rho^2$.

In order to find a phase matching relation, we expand the dispersion relation for the perturbations $k(\omega)$ for $|\omega|\gg 2\sqrt{\gamma P/\beta_2}$
\begin{equation}\label{k}
k=\sqrt{\frac{\beta_2\omega^2}{2}\left(\frac{\beta_2\omega^2}{2}+2\gamma P\right)}\approx\frac{\beta_2\omega^2}{2}+\gamma P.
\end{equation}
In this way we have
$$\tilde \Delta\approx 2\rho\cos[kL+\psi_\text{e}(\omega)+\phi].$$
The potentially unstable frequencies maximise  $|\tilde\Delta|$, and thus satisfy the following equation:
\begin{equation}\label{PM}
k(\omega)L+\phi+\psi_\text{e}(\omega)=m\pi,\;m=0,\pm 1,\ldots
\end{equation} 

The solutions of  Eq. (\ref{PM}) for $m\ne 0$ correspond to parametric resonances (PRs) induced by the periodic forcing represented by the injection of the pump at each roundtrip \cite{Conforti2016}. 
We concentrate on the $m=0$ band 
and use the expansion (\ref{k}), to get the following simple phase-matching relation:
\begin{equation}\label{PM1}
\frac{\beta_2\omega^2}{2}L+2\gamma PL+\phi_0+\psi_\text{e}(\omega)=0
\end{equation}
Equation (\ref{PM1}) has a straightforward physical meaning: the phase acquired by the perturbations propagating on top of the cw ($\beta_2\omega^2L/2 +\gamma P$) plus the total phase shift of the cavity (linear+nonlinear: $\phi=\phi_0+\gamma P$) plus the even part of the phase of the filter ($\psi_e(\omega)$) must be zero to have parametric amplification. Equation (\ref{PM1}) is a generalisation Eq.(8) of Ref. \cite{Coen1997}, including the dispersion induced by the filter.

\section{The mean field model}
\subsection{Averaging the Ikeda map}
In this section, we derive a mean field model, i.e. a generalized Lugiato-Lefever equation, by performing a suitable averaging of the Ikeda map Eqs. (\ref{nls},\ref{newBC}).
Using Eq. (\ref{nls}), we can formally approximate at first order the field envelope $A_n$ at spatial position $L$ after propagation from $z=0$ to $z=L$ as follows:
\begin{eqnarray}
&&A_{n}(L,t)\approx A_{n}(0,t)+L\left . \frac{\partial A_{n}(z,t)}{\partial z}\right |_{z=0}=\\ \nonumber
&&A_{n}(0,t)+\left[\frac{-iL\beta_2}{2}\frac{\partial^2}{\partial t^2}+iL\gamma|A_{n}(0,t)|^2\right]A_{n}(0,t).
\end{eqnarray}
Assuming that filter and coupler are located at the same position $z=L$, the Fourier transform of the field, $\hat{A}(\omega,0)$, obeys the  boundary conditions  described by Eq. (\ref{newBCfreq}) :
\begin{eqnarray}\label{bcf2}
\nonumber\hat{A}_{n+1}(0,\omega)=\theta E_{IN}\delta(\omega)+\rho e^{i\phi_0}e^{F(\omega)+i\psi(\omega)}\hat{A}_{n}(L,\omega),\\
\end{eqnarray}
where we have explicitly written $H(\omega)=e^{F(\omega)+i\psi(\omega)}$.

 
We assume $\rho\approx 1$, $\theta\ll1$, $|\phi_0|\ll 1$ , $F,|\psi|\ll1$, and we expand in Taylor series at the first order all the terms in the boundary conditions Eq. (\ref{bcf2}). We neglect all the products corresponding to different physical effects, and after taking the inverse Fourier transform we obtain:
\begin{eqnarray}
A_{n+1}-A_{n}&=&\left[-\alpha+i\phi_0+\Phi\star+i\Psi\star\right]A_{n}+\\\nonumber
&+&\left[\frac{-iL\beta_2}{2}\frac{\partial^2}{\partial t^2}+iL\gamma|A_{n}|^2\right]A_{n}+\theta\sqrt{P_{IN}}.
\end{eqnarray}
%
where we used the notation $A_{n}=A_{n}(t,0)$. Note that $\Phi$ and $\Psi$ are the inverse Fourier transforms of $F$ and of $\psi$ respectively. 
 By approximating the spatial derivative with the difference quotient $(A_{n+1}-A_n)/L\approx \partial A/\partial z|_{z=nL}$,
we can pass the map to the continuous limit obtaining the following equation for the field $A(z,t)$:
\begin{eqnarray}\label{GLLE}
L\frac{\partial A}{\partial z}&=&\left[-\alpha+i\phi_0+\Phi\star+i\Psi\star\right]A+\\ \nonumber
&+&\left[\frac{-iL\beta_2}{2}\frac{\partial^2}{\partial t^2}+iL\gamma|A|^2\right]A+\theta \sqrt{P_{IN}}.
\end{eqnarray}
 Equation (\ref{GLLE}) represents a mean-field generalised Lugiato-Lefever equation, which differs from the standard  LLE \cite{LLE1,LLE2,Haelterman92} by the presence of filter terms.

\subsection{Linear stability analysis}
Equation (\ref{GLLE}) admits a continuous wave homogeneous solution with power $\bar{P}$ which is determined by the characteristic bistable response of the resonator according to the following relation:
\begin{eqnarray}
\bar{P}=\frac{\theta^2}{\left(-\alpha+F(0)\right)^2+\left(\phi_0+\psi(0)+\gamma L\bar{P}\right)^2}P_{IN}.
\end{eqnarray}
We perform a linear stability analysis of the CW solution by inserting the following \emph{ansatz}
\begin{eqnarray}
A(z,t)=A_0+A_+(z)e^{-i\omega t}+A_-(z)e^{i\omega t},
\end{eqnarray}
into Eq. (\ref{GLLE}) where $A_0=\sqrt{\bar{P}}e^{i\xi}$, being $\xi$ a phase factor, and $A_+$, $A_-$ the amplitudes of perturbations oscillating at frequency detuned by $\mp\omega$ with respect to the CW solution. Linearising with respect to the small perturbations ($|A_0|>>|A_-|,|A_+|$) we obtain the following system of coupled equations:


\begin{eqnarray}\nonumber
L\frac{\partial A_+}{\partial z}&=&iL\omega^2\frac{\beta_2}{2}A_++i\phi_0A_++i\psi(\omega)A_++F(\omega)A_++\\&+&i\gamma L2\bar{P}A_++i\gamma L \bar{P}e^{2i\xi}A_-^*-\alpha A_+\\ \nonumber
L\frac{\partial A_-^*}{\partial z}&=&-iL\omega^2\frac{\beta_2}{2}A_-^*-i\phi_0A_-^*-i\psi(-\omega)A_-^*+F(-\omega)A_-^*\\&-&i\gamma L2\bar{P}A_-^*-i\gamma L \bar{P}e^{-2i\xi}A_+-\alpha A_-^*.
\end{eqnarray}
A phase rotation and amplitude rescaling allows us to get a better insight on how the filter acts on the perturbations and hence to better appreciate the contributions to MI. We hence perform the following change of variables: 
\begin{eqnarray}
A_+&=&a_+e^{[i\psi_{o}(\omega)+F_e(\omega)]z-\alpha z}\\
A_-^*&=&a_-^*e^{[i\psi_{o}(\omega)+F_e(\omega)]z-\alpha z}
\end{eqnarray}
which leads to
\begin{eqnarray}\nonumber
L\frac{\partial a_+}{\partial z}&=&iL\omega^2\frac{\beta_2}{2}a_++i\phi_0a_++i\psi_{e}(\omega)a_++F_o(\omega)a_++\\\label{s1}&&+i\gamma L2\bar{P}a_++i\gamma L \bar{P}e^{2i\xi}a_-^*\\ \nonumber
L\frac{\partial a_-^*}{\partial z}&=&-iL\omega^2\frac{\beta_2}{2}a_-^*-i\phi_0a_-^*-i\psi_{e}(\omega)a_-^*-F_o(\omega)a_-^*-\\\label{s2} &&i\gamma L2\bar{P}a_-^*-i\gamma L \bar{P}e^{-2i\xi}a_+
\end{eqnarray}
where the even and odd parts of $F(\omega)$ have been defined as:
\begin{align}
F_\text{e}(\omega)=\frac{F(\omega)+F(-\omega)}{2},\\
F_\text{o}(\omega)=\frac{F(\omega)-F(-\omega)}{2}.
\end{align}
We can now easily recast the system evolution in a matrix form:
\begin{eqnarray}
L\frac{\partial}{\partial z}
\begin{pmatrix} 
a_+ \\
a_-^*
\end{pmatrix}
=M
\begin{pmatrix} 
a_+ \\
a_-^*
\end{pmatrix}
\end{eqnarray}

where the evolution matrix reads 
\begin{eqnarray}\nonumber
M=\begin{pmatrix} 
i\mu+F_o(\omega) & i\gamma L \bar{P}e^{2i\xi} \\
-i\gamma L \bar{P}e^{-2i\xi} &-i\mu-F_o(\omega)
\end{pmatrix}
\end{eqnarray}
with 
\begin{eqnarray}\label{Delta}
\mu=L\omega^2\frac{\beta_2}{2}+2\gamma\bar{P} L+\phi_0+\psi_{e}(\omega)
\end{eqnarray}
a phase-mismatch parameter. It is worth noting that Eq. (\ref{Delta}) is equivalent to the phase-matching condition Eq. (\ref{PM1}) obtained from the linear stability analysis of the Ikeda map developed in the previous sections.
The eigenvalues of the matrix $M$ read:
 \begin{eqnarray}
\lambda_\pm=\pm\sqrt{-\left[\mu(\omega)-iF_o(\omega)\right]^2+(\gamma L \bar{P})^2},
 \end{eqnarray}
thus we define the MI gain as:
  \begin{eqnarray}\label{gainLLE}
g_{LLE}(\omega)=2\frac{-\alpha+F_e(\omega)+ Re(\lambda_+)}{L}.
  \end{eqnarray}
 It follows that the power of the perturbations $|A_\pm|^2$ grows exponentially as $\exp[g_{LLE}(\omega)z]$ when $g_{LLE}(\omega)>0$.  
  

Starting from Eqs. (\ref{s1}) and (\ref{s2}) we can obtain a physical insight into the meaning of the mismatch parameter $\mu$ defined in Eq. (\ref{Delta}). By neglecting the dissipative part of the filter, Eqs. (\ref{s1}) and (\ref{s2}) can be recast in the following form:
\begin{eqnarray}\label{m1}
L\frac{\partial a_+}{\partial z}&=&i\gamma L \bar{P}e^{2i\xi}a_-^*e^{-2i\mu z}\\\label{m2}
L\frac{\partial a_-^*}{\partial z}&=&-i\gamma L \bar{P}e^{2i\xi}a_+e^{2i\mu z}.
 \end{eqnarray}

From Eqs. (\ref{m1}) and (\ref{m2}) we can see that at phase-matching, when $\mu=0$, $a_+$ and $a_-^*$ grow exponentially in $z$. It is apparent that the standard cavity mismatch can be compensated by the presence of the filter even in normal dispersion and for zero or positive detuning for the pump.


\section{control of the instability gain}
In this section we analyse the effect of several control parameters on the instability gain. We present a systematic comparison of the results of the stability analysis from the Ikeda map and the generalised LLE, which permits to appreciate the accuracy and the limits of the mean-field model.
Even if our results are general, in this section we restrict to the normal dispersion $\beta_2>0$ and monostable regime $\delta/\alpha<\sqrt3$. In this case the standard Turing instability can not develop \cite{Haelterman92}  and the instability is induced only by presence of the filter.
\begin{figure}[b]
    \centering
    \includegraphics[width=0.45\textwidth]{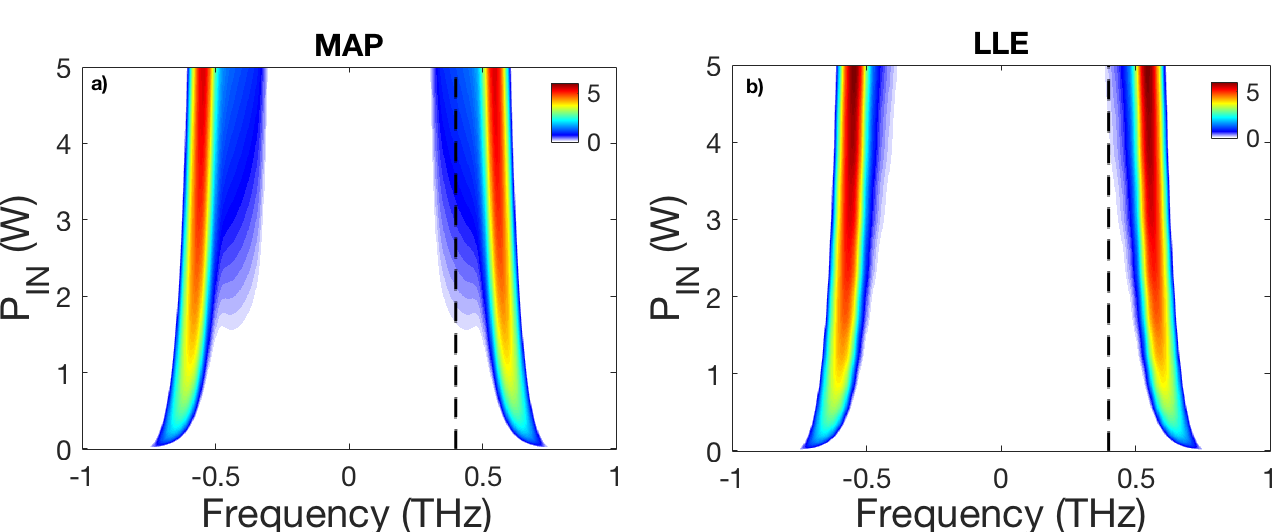}
    \caption{The positive part of the MI gain is depicted a function of $P_{IN}$, calculated from a) the map and b) the LLE. The dashed lines denote filter position $\omega_f/(2\pi)$. Parameters used are: $\beta_2=0.5$ ps$^{2}$km$^{-1}$, $\gamma=2.5$ W$^{-1}$km$^{-1}$, $L=0.1$ km, $\rho=\sqrt{0.9}$, $\phi_0=-\psi(0)$, $\theta=\sqrt{0.1}$, $a=500$ rad/ns, $b$=-3.2, $\omega_f=2\pi\cdot400$ rad/ns.}
    \label{figure2}
\end{figure}
In Fig.\ref{figure2}a) we show the MI gain as a function of the pump power calculated for the Ikeda map [see Eq. (\ref{gain_map})]. We can note the presence of an unstable band at a frequency slightly higher than the central position of the filter (dashed black line), and of its symmetric at negative frequency shift. The gain increases monotonically with input power, while a decreasing trend of the maximally unstable frequency is observed. Fig.\ref{figure2}b) reports the gain obtained from the analysis of LLE [see Eq. (\ref{gainLLE})]. The mean-field reproduces qualitatively the same picture. In particular, we note a quantitative agreement regarding the peak spectral position and amplitude of the unstable bands. For high input powers (greater than $\approx 2$ W), the map predicts a small additional lobe peaked at the specral position of the filter, which is not captured by LLE.


\begin{figure}[b]
    \centering
    \includegraphics[width=0.5\textwidth]{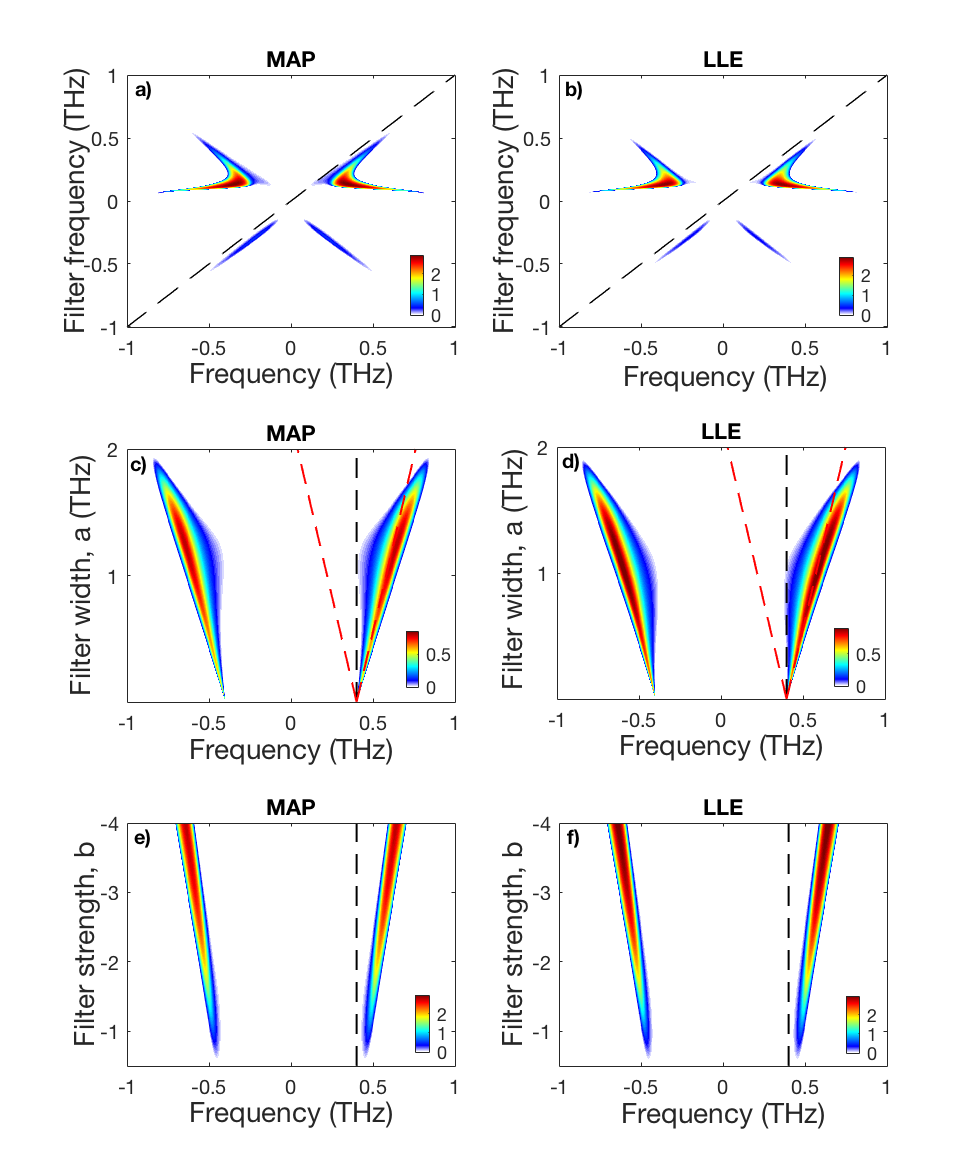}
    \caption{The positive part of the MI gain is depicted for map and LLE as a function of the filter frequency shift with respect to the pump $\omega_f/(2\pi)$ (panels a) and b)), of the filter width $a$ (panels c) and d)); and of the filter strength $b$ (panels e) and f)). The dashed black lines denote filter position, whereas the dashed red lines denote the filter width at half maximum. Parameters used are: $\beta_2=0.5$ ps$^{2}$km$^{-1}$, $P_{IN}=0.5$ W, $\gamma=2.5$ W$^{-1}$km$^{-1}$, $L=0.1$ km, $\rho=\sqrt{0.95}$, $\phi_0=-\psi(0)$, $\theta=\sqrt{0.05}$, $a=400$ rad/ns in a), b), e) and f); $b$=-1 in a), b), c) and d); $\omega_f=2\pi\cdot 400$ rad/ns in c), d), e) and f).}
    \label{figure3}
\end{figure}
The dependency of the MI gain on the three filter parameters, namely frequency shift (with respect to the pump) $\omega_f$, spectral width $a$ and strength $b$ is shown in Fig.\ref{figure3}. Figure \ref{figure3}a) reports the gain as a function of $\omega_f$ calculated from the map.  The band is located at a slightly higher frequency than the filter, except when the filter is very close to the pump, where we observe a shift of the gain band towards higher frequencies. This happens because for this analysis we decided to compensate the filter induced phase shift  with the cavity phase shift $\phi_0=-\psi(0)$ in order to stay in the monostable regime, and the filter phase profile has a slowly decreasing tail (see Fig. \ref{figure1}a)).  We also note that for positive frequency shift $\omega_f$ the gain is substantially higher. The reason for this asymmetry is that, unlike the amplitude $F$, the filter phase $\psi$ is an odd function (with respect to the central frequency $\omega_f$), so it acts in a substantially different way depending if it is placed at positive or negative frequency shift with respect to the pump. The results obtained from the mean-field model shown in Figure \ref{figure3} b) are practically identical.
Figure \ref{figure3}c) reports the gain as a function of the filter parameter $a$, which mainly controls the filter with, for a fixed filter frequency shift $\omega_f/(2\pi)=400$ GHz calculated from the map. The unstable band is located at the high-frequency edge of the filter, which can be calculated as $\omega_f+\Delta\omega_{HWHM}$ (see definition after Eq. (\ref{phaseF})). The peak gain increases with $a$ and reaches a maximum for $a\approx1.2$ THz. Above this value the peak gain decreases, to eventually vanish for values greater than $\approx 2$ THz. This drop in gain takes place because the filter start to cut the pump, reducing the intracavity power (input power is fixed here), which eventually controls the parametric gain. Also in this case, the results obtained from the mean-field model shown Figure \ref{figure3} d) are in perfect agreement.
In Fig. \ref{figure3}e) we show the gain from the map as a function of the parameter $b$, which determines the maximum attenuation of the filter, but also the amplitude of the phase response. For shallow filters $-0.2 \lesssim b<0$ the system is modulationally stable. By increasing the strength of the filter, a band appears at a frequency greater than the filter central frequency $\omega_f/(2\pi)$ and whose peak amplitude and spectral position grows monotonically with $|b|$. This evolution is ruled by the phase profile of the filter which increases with $|b|$ and shifts the phase-matching frequency towards higher values, as shown by Eq. (\ref{PM1}). Quite surprisingly, the mean field approach still works perfectly (see \ref{figure3}f)), even when the perturbation induced by the filter at each roundtrip are not at all small. This is a further confirmation that LLE hold valid beyond the assumptions traditionally used for its derivation \cite{Conforti2014,Copie2017,Conforti2017,Bessin2019}.


\begin{figure}[htb]
    \centering
    \includegraphics[width=0.5\textwidth]{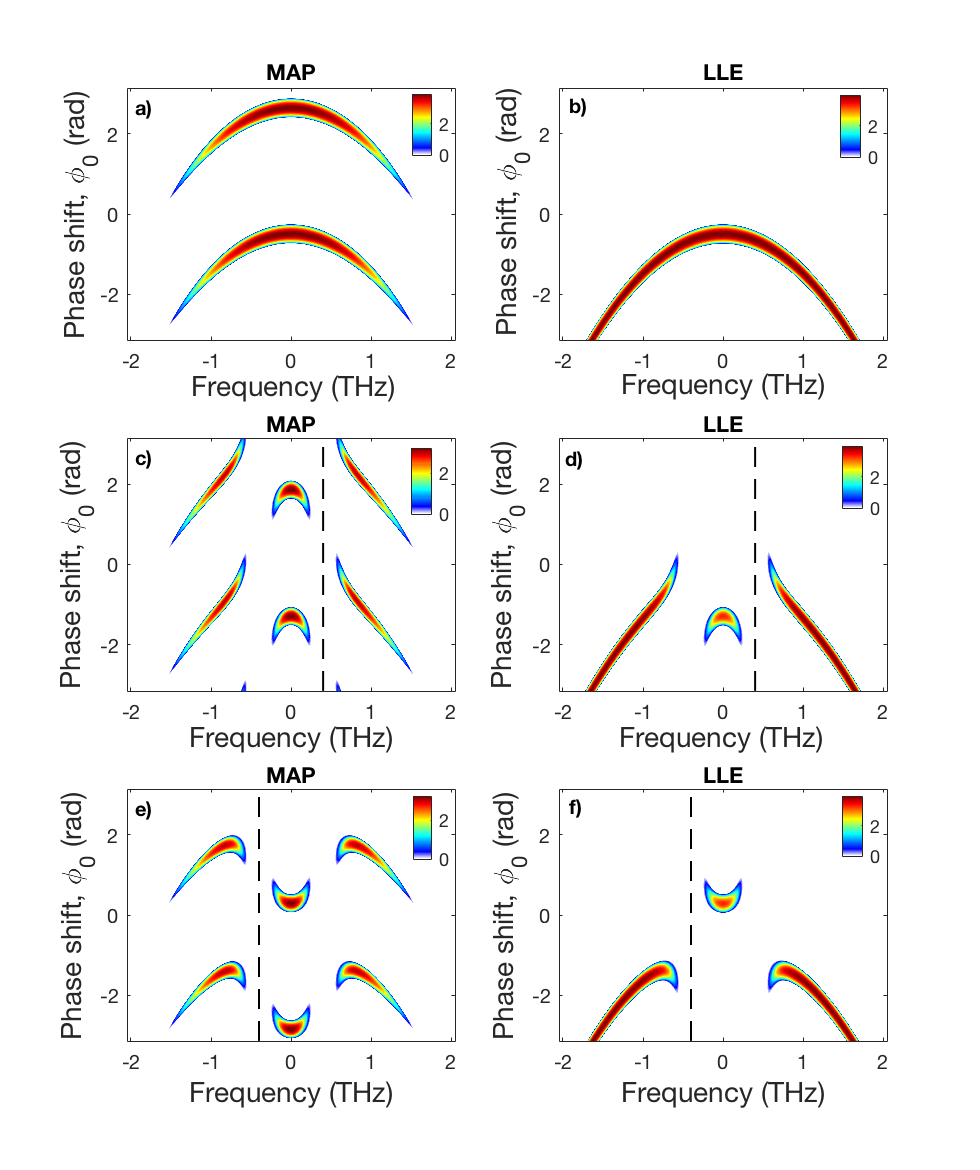}
    \caption{The positive part of the MI gain is depicted as a function of the phase shift $\phi_0$ for the map without filter a), with filter positively detuned with respect to the pump ($\omega_f/(2\pi)=400$ GHz) c), and with filter negatively detuned with respect to the pump ($\omega_f/(2\pi)=-400$ GHz) e); the corresponding LLE cases are plotted in b), d) and e). The dashed lines denote filter position. Parameters used are: $\beta_2=0.5$ ps$^{2}$km$^{-1}$, $\gamma=2.5$ W$^{-1}$km$^{-1}$, $L=0.1$ km, $\rho=\sqrt{0.9}$, $\theta=\sqrt{0.1}$,  $a=800$ rad/ns, $b$=-3.2 and intracavity power 1 W.}
    \label{figure4}
\end{figure}

The study of the dependence of the MI gain on the cavity phase shift is presented in Fig.\ref{figure4}. Figures  \ref{figure4} a) and b) show the MI gain for the map and the LLE, calculated in absence of the filter. For the map, two branches are present: the lower branch corresponds to the Turing instability, which is maximal near the cavity resonance ($\phi_0=0$) and is also captured by the LLE. The upper one corresponds to the parametric resonance induced by the periodic boundary conditions, is peculiar to the map, and develops patterns which are in anti-phase roundtrip after roundtrip (also called period 2 or P2 MI) \cite{Haelterman92,Coen1997,Conforti2016}.
Figures \ref{figure4} c)-f) demonstrate a strong modification of the instability spectrum described above induced by the filter presence. Each branch is split around the spectral position of the filter. The low-frequency part  is moved towards lower (higher) phase-shifts and the edges of the high-frequency parts are bent upwards (downwards) when the filter central position is $\omega_f/(2\pi)=400$ GHz (-400 GHz). Hence the gain spectrum reveals a strong asymmetry depending on whether the filter frequency is positively or negatively detuned with respect to the pump.  This suggests a further degree of freedom for parametric gain engineering and control in driven optical cavities. 

\begin{figure}[htb]
    \centering
    \includegraphics[width=0.5\textwidth]{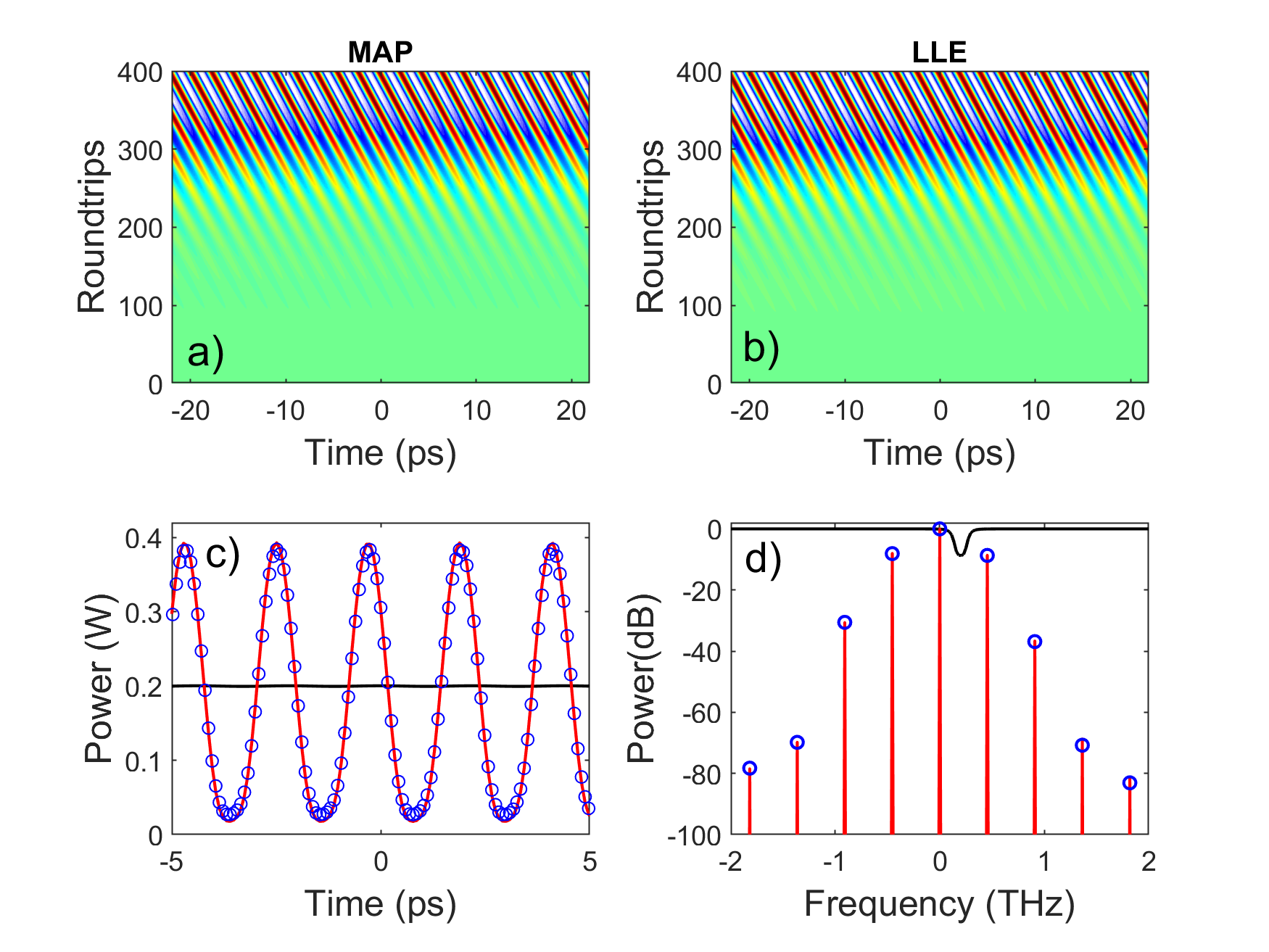}
    \caption{
    The evolution of the field power profile over roundtrips from a) the Ikeda map and b) the LLE.
    The output power pattern and the corresponding power spectrum (normalised to its maximum) are shown  in panels c), d). Solid red curves denotes results from the Ikeda map and blue circles from the LLE. The solid black curve in c) represents the initial condition, whereas in d) it represent the filter amplitude $|H(\omega)|$. The parameters are 
    $\beta_2=0.5$ ps$^{2}$km$^{-1}$, $\gamma=2.5$ W$^{-1}$km$^{-1}$, $L=0.1$ km, $\rho=\sqrt{0.95}$, $\phi_0=-\psi(0)$, $\theta=\sqrt{0.05}$, $a=400$ rad/ns, $b=-1$, $\omega_f/(2\pi)=200$ GHz,
    $P_{IN}=0.0149$ W for the Ikeda map and $P_{IN}=0.015$ W for the LLE (intracavity power of stationary solution is 0.2 W in both cases).
    Simulations initial conditions were $A(0,t)=\sqrt{0.2}[1+0.001\cos(2\pi\nu_{max}t)]$ ($\nu_{max}=455 GHz$).}
    \label{figure5}
\end{figure}
%

\section{Frequency combs and temporal patterns}
The presence of modulationally unstable frequencies in dissipative systems can lead to the generation of periodic trains of pulses, which correspond to frequency combs in the spectral domain \cite{Bessin}. The generated pulse train may be interpreted as a stable attractor of the infinite dimensional dissipative system described Eq. (\ref{GLLE})  \cite{Haelterman92}. This behaviour contrasts the recurrence phenomenon observed in the NLSE \cite{VanSimaeys2001,Vanderhaegen2020}, which is a conservative Hamiltonian system.  In Fig. \ref{figure5}  we report an example of the generation of a stable temporal pattern using a shallow filter blue-detuned with respect to the pump wavelength. Figure \ref{figure5}a) reports the power evolution simulated with the Ikeda map of an initial condition consisting in the cw solution with a small sinusoidal perturbation (see figure caption for more details). After the initial stage of perturbation growth, a stable pattern is generated around the 300th roundtrip. The temporal drift towards negative delay is caused by the odd part of the filter phase, which act as an additional dispersion. The evolution simulated from LLE is reported in Fig. \ref{figure5}b) is practically identical, showing once again the accuracy of the mean-field model.
The output pattern at roundtrip 400 is plotted in temporal and spectral domain in Figs. \ref{figure5}c,d), where red curves, respectively blue dots stands for Ikeda map and LLE, respectively.  The comparison of the time domain pattern and power spectra demonstrate excellent quantitative agreement between map and LLE nonlinear solutions for parameters consistent with the assumptions made in the mean field model derivation. We have also verified that at much higher pump power the Ikeda map exhibits a richer dynamics as it can be naturally expected due to its broader validity range in parameters space.

\section{Conclusions}
In this article we have presented the theory of filter induced modulation instability in passive driven Kerr cavities. Starting from an Ikeda map model we have derived a generalized mean field equation of the Lugiato-Lefever type. We have performed a linear stability analysis of the homogeneous solutions of both models and our results show the existence of a peculiar kind of modulation instability developing also in normal group velocity dispersion and in Turing-stable regimes. Besides agreeing well in their theoretical predictions both the mean field model and the map describe equally well the nonlinear stage of the filter induced MI, consisting in the generation of stable frequency combs. We have specialised our analysis on a particular shape of minumum-phase filter, which is quite general for the modelling of fiber Bragg gratings. The theory can be straightforwardly generalised to other filter responses more suited to other kinds of resonators such as Fabry-Perot cavities or microresonators.
The results presented in this article will be relevant for the the development and design of novel sources of coherent light with tuneable features.
Indeed controlling the frequency shift between the pump the filter would allow tuneability of the frequency position of the generated spectral sidebands and ultimately the possibility of tuneable frequency comb generation in the nonlinear stage of the filter induced MI as pioneered in \cite{Bessin}.

\section{Acknowledgements}
This study was partly supported by IRCICA, by the French government through the Programme Investissement d’Avenir (I-SITE ULNE / ANR-16-IDEX-0004 ULNE, projects VERIFICO, EXAT, FUNHK) managed by the Agence Nationale de la Recherche.
The work of A.M.P. was supported by the Royal Academy of Engineering under the Research Fellowship scheme.


\begin{thebibliography}{52}
\bibitem [{()}]{notehilbert}%
  \BibitemOpen
  \href@noop {} {}\bibinfo {note} {Note that an alternative definition of the
  Hilbert transform exists which has a minus sign in front of the integral
  \cite{Bracewell}. If this alternative definition is used, the sign of the
  phase in Eq. (\ref{Bode}) is changed, as in \cite{Bessin}}\BibitemShut
  {NoStop}%
\bibitem [{not()}]{note_delay}%
  \BibitemOpen
  \href@noop {} {}\bibinfo {note} {A linear spectral phase $e^{i\omega t_0}$
  equivalent to a delay $t_0$ in time domain is not a minimum-phase response,
  because the exponential has an essential singularity at $|\omega|\rightarrow
  \infty$ and the logarithm is not analytic. Even if the filter response is
  likely to contain a linear spectral phase in the laboratory reference frame,
  corresponding to propagation through the length of the filter, it disappears
  in the pulse reference frame ($t=t_{lab}-z/v_g$) used in our formulation of
  NLSE Eq. (\ref{nls}).}\BibitemShut {Stop}%
\end{thebibliography}
%

\end{document}